\documentclass[final,num-names,3p,times]{elsarticle}

\usepackage{graphicx}
\usepackage{amsmath, amssymb, amsthm}
\usepackage{cases}
\usepackage{subfigure}
\usepackage{multirow}

\biboptions{sort}

\journal{Discrete Applied Mathematics}

\newtheorem{thm}{Theorem}
\newtheorem{cor}{Corollary}

\def\para#1{\paragraph{\bf #1}} 
\def\para#1{\subsection{#1}}

\def\lp{\hbox{\tt (}}
\def\rp{\hbox{\tt )}}
\def\yes{\checkmark}
\def\no{$\times$}

\begin{document}

\begin{frontmatter}
\title{Complexity of the path avoiding forbidden pairs problem revisited}
\author{Jakub Kov\'a\v c}
\ead{kuko@ksp.sk}
\address{Department of Computer Science,
           Comenius University,
           Mlynsk\'a Dolina,\\842~48 Bratislava, Slovakia}

%
%

\begin{abstract}
Let $G=(V,E)$ be a directed acyclic graph with two distinguished vertices $s, t$,
and let $F$ be a set of \emph{forbidden pairs} of vertices. We say that a path in $G$
is \emph{safe}, if it contains at most one vertex from each pair $\{u,v\}\in F$.
Given $G$ and $F$, the \emph{path avoiding forbidden pairs} (PAFP) problem is
to find a safe $s$--$t$ path in $G$.

We systematically study the complexity of different special cases of the PAFP problem
defined by the mutual positions of fobidden pairs.
Fix one topological ordering $\prec$ of vertices; we say that pairs $\{u,v\}$ and $\{x,y\}$ are
\emph{disjoint}, if $u\prec v \prec x\prec y$, \emph{nested}, if $u\prec x\prec y \prec v$, and
\emph{halving}, if $u\prec x\prec v\prec y$.

The PAFP problem is known to be NP-hard in general or if no two pairs are disjoint;
we prove that it remains NP-hard even when no two forbidden pairs are nested.
On the other hand, if no two pairs are halving, the problem is known to be solvable in
cubic time. We simplify and improve this result by showing an $O(M(n))$ time algorithm,
where $M(n)$ is the time to multiply two $n\times n$ boolean matrices.
\end{abstract}
\begin{keyword}
path \sep forbidden pair \sep NP-hard \sep dynamic programming
\end{keyword}
\end{frontmatter}

\section{Introduction}

Let $G=(V,E)$ be a directed graph with two distinguished vertices $s, t\in V$ and
let $F\subseteq {V\choose 2}$ be a set of \emph{forbidden pairs} of vertices.
We say that a path $\pi$ is \emph{safe}, if it does not contain any forbidden pair,
i.e., $\pi$ contains at most one vertex from each pair $\{u,v\}\in F$.
Given $G$ and $F$, the \emph{path avoiding forbidden pairs} problem (henceforth PAFP)
is to find a safe $s$--$t$ path in $G$.
In this paper, we study the complexity of different special cases 
of the problem on directed \emph{acyclic} graphs.

\para{Motivation}
The PAFP problem was first studied by \citet{krause} and \citet{srimani}
motivated by designing test cases for automatic software testing and validation.
We can represent a program as a directed graph where vertices represent segments
of code and edges represent the flow of control from one code segment into another.
The goal is to cover this graph with $s$--$t$ paths corresponding to different test cases.
However, not all paths correspond to executable sequences in the program.
Therefore \citet{krause} introduced forbidden pairs which identify the 
mutually exclusive code segments 
and formulated the PAFP problem. Unfortunatelly, as shown by \citet{gabow},
the problem is NP-hard even for directed acyclic graphs.

A different motivation came from bioinformatics and the problem of peptide sequencing
via tandem mass spec\-tro\-met\-ry. Peptides are polymers which
can be though of as strings over a 20 character alphabet of amino acids and
the sequencing problem is to determine the amino acid sequence of a given peptide.
To this end, many copies of the peptide are fragmented and the mass
of the fragments is measured (very precisely) by a mass spectrometer. 
The result of the experiment is a mass spectrum where
each peak corresponds to mass of some prefix or some suffix of
the amino acid sequence, or is a noise. The spectrum is then compared
against a database of known fragment weights.

\citet{chen} suggested the following formulation of the peptide sequencing problem:
Let us create a \emph{spectrum graph} with two vertices $p_i$ and $s_i$
for each peak $w_i$ with weights $w(p_i)=w_i-1$ and $w(s_i)=W-w_i+1$, where $W$
is the weight of the whole peptide.
We add an edge from $x$ to $y$ if the difference between
weights $w(y)-w(x)$ equals the total mass of some known sequence of amino acids.
Thus, paths in this graph correspond to amino acid sequences.
Paths going through $p_i$ correspond to $w_i$ being a weight of some prefix
and similarly, paths going through $s_i$ correspond to $w_i$ being a weight
of some suffix. (Paths going through neither $p_i$ nor $s_i$ correspond to $w_i$ being a noise.)
However, $w_i$ cannot be a prefix weight \emph{and} a suffix weight
at the same time, so $\{p_i,s_i\}$ will form a forbidden pair for each $i$.
This is a very special case of the PAFP problem in directed acyclic graphs 
where all the forbidden pairs are nested and \citet{chen} showed that it is
polynomially solvable.

The PAFP problem on directed acyclic graphs also arose in a completely different application
in bioinformatics -- gene finding using RT-PCR tests \citep{pcr}.
In this application, we have a so called \emph{splicing graph} where vertices
represent non-overlapping segments of the DNA sequence, length of a vertex is
the number of nucleotides in this segment, and edge $(u, v)$ indicates
that segment $v$ immediately follows segment $u$ in some gene transcript.
Thus, paths in this splicing graph correspond to putative genes.
The problem is to identify the true genes with a help of information from RT-PCR experiments.

Without going into biology details, let us define a (simplified) result of an RT-PCR experiment
as a triple $t=(u,v,\ell)$, where $u,v\in V$ are two vertices and $\ell$ is the length of a product.
Let $\pi$ be a path going through $u$ and $v$ in the splicing graph; if the length of the $u$--$v$
subpath is equal to $\ell$, we say that $\pi$ \emph{explains} test $t$, otherwise, it is \emph{inconsistent}
with test $t$. We can define a score of a path $\pi$ with respect to a set of tests $T$ as a sum of
the scores of all of its vertices and edges, plus a bonus $B$ for each explained test from $T$, and
minus a penalty $P$ for each inconsistent test. The \emph{gene finding with RT-PCR tests} problem
is to find an $s$--$t$ path with the highest score in the given splicing graph $G$ with a set of RT-PCR tests $T$.

Note that if we set all lengths to an unattainable value, say $-1$, and we set a high (infinite) penalty $P$
for inconsistent tests, we basically get the PAFP problem. Thus, the PAFP problem is at the core of gene
finding with RT-PCR tests and the latter problem inherits all NP-hardness results for the PAFP problem.
On the positive side, we have shown in our previous work \citep{pcr} that some polynomial solutions
for special cases of the PAFP problem can be extended to pseudo-polynomial algorithms for the gene finding problem.

\para{Previous results}

As shown by \citet{gabow}, the PAFP problem is NP-hard in general, but several special cases
are polynomially solvable. \citet{yinnone} studied the PAFP problem under \emph{skew symmetry}
conditions where for each two forbidden pairs $\{u,u'\}, \{v, v'\}\in F$, \emph{if} there is an
edge from $u$ to $v$, there is also an edge from $v'$ to $u'$. He proved that under such conditions,
the problem is polynomially equivalent to finding an augmenting path with respect to a given matching
and thus polynomially solvable. 

For directed acyclic graphs, we have already mentioned that the nested case is solvable in polynomial
time \citep{chen}; \citet{kp} were able to devise a polynomial algorithm if the set of forbidden pairs
has a well-parenthesized or a halving structure (see Preliminaries).

Recently, approximability and parameterized complexity of the PAFP problem have been studied:
We add 1 to the objective function to disallow a zero cost solutions -- otherwise the problem is
trivially inapproximable. \citet{checkpoint} showed that even then there is a constant $c>0$ such
that minimizing $1+\null$the number of forbidden pairs on an $s$--$t$ path is not $c\cdot n$-approximable. 
\citet{w1} studied the PAFP problem on undirected graphs. When parameterized by the vertex cover of $G=(V,E)$,
the problem is W[1]-hard (the proof also carries over to directed acyclic graphs). On the other hand,
when parameterized by the vertex cover of $H=(V, F)$ (where edges are forbidden pairs), the problem is
fixed parameter tractable (FPT), but has no polynomial kernel unless $\hbox{NP}\subseteq \hbox{coNP}/\hbox{poly}$.
The problem is also FPT when parameterized by the treewidth of $G\cup H$.


\para{Contributions and road map}

In this paper, we systematically study different special cases of the PAFP problem on topologically sorted directed acyclic graphs.
In the next section, we introduce the different special cases based on mutual positions of forbidden pairs.
In Section~\ref{s:ordered}, we prove that the PAFP problem is NP-hard even if the set of forbidden pairs
has ordered structure and in Sections~\ref{s:paren} and \ref{s:concl}, we improve upon the results of \citet{chen} and \citet{kp}
for the nested, halving, and well-parenthesized forbidden pairs.

\section{Preliminaries} \label{s:prelim}

Let $G=(V,E)$ be a directed acyclic graph and let $F$ be the set of forbidden pairs.
In this work, we assume that $G$ is topologically sorted and the linear order of
vertices is given. This is less general than in the work of \citet{kp}, on the other
hand, it is well motivated by applications -- all the graphs mentioned
in the Motivation section have natural linear ordering.
We say that vertex $u$ \emph{is before} or \emph{precedes} $v$,
$u\prec v$, if $u$ precedes $v$ in this linear order.

As already noticed by \citet{yinnone} and \citet{kp}, we may assume that 
no vertex belongs to more than one forbidden pair. If this is the case, we can
replace a vertex with $k>1$ forbidden pairs by a directed path of length $k$
and move each end of a forbidden pair to a different vertex on this path.
This preprocessing can be done in linear time and the number of vertices and edges
is increased only by $|F|$ at most.

To define special cases of interest, let us denote the forbidden pairs $\{f_i,f_i'\}$
for $i=1,\ldots,k$, where $f_i\prec f_i'$ and $f_1\prec f_2\prec\cdots\prec f_k$,
i.e., we order them by position of the left member of the pair.
We recognize three possible types of mutual position of pairs $\{u,v\}$
and $\{x,y\}$ (without loss of generality, let $u\prec v$, $x\prec y$, and $u\prec x$):
\emph{disjoint} ($u,v \prec x,y$; see Fig.~\ref{fig:disjoint-pair}),
\emph{nested} ($u\prec x,y \prec v$; see Fig.~\ref{fig:nested-pair}), and
\emph{halving} ($u\prec x\prec v\prec y$; see Fig.~\ref{fig:halving-pair}).
All the special cases are obtained by restricting the set of forbidden pairs $F$
to only certain types of mutual positions (see Table~\ref{tab:pafp}).
This gives us $2^3=8$ cases, from which these 6 classes are non-trivial and interesting:

\begin{figure}[htp]
  \centering
  \subfigure[disjoint pairs]{ \includegraphics[scale=0.7]{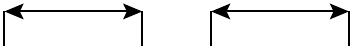}\label{fig:disjoint-pair} }\qquad
  \subfigure[nested pairs]{  \includegraphics[scale=0.7]{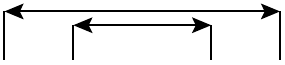}\label{fig:nested-pair} }\qquad
  \subfigure[halving pairs]{ \includegraphics[scale=0.7]{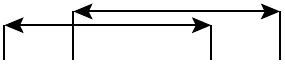}\label{fig:halving-pair}}
  \vskip-8pt
  \caption{Different mutual positions of two forbidden pairs.}
\end{figure}

\begin{table*}
\centering
\caption{Complexity of the PAFP problem for its different special cases;
  $n$ and $m$ denote the number of vertices and edges of $G$, respectively; $O(n^\omega)$
  is the complexity of boolean matrix multiplication, $\omega< 2.3727$ \citep{mm,mm2}.}
  \label{tab:pafp}
\smallskip
\noindent\makebox[\textwidth]{
\begin{tabular}{cccccc}
\multirow{2}{*}{\sc{Problem}} & \multicolumn{3}{c}{\sc Allowed Forbidden Pairs} & \multirow{2}{*}{\sc{Complexity}} & \multirow{2}{*}{\sc{Example}} \\
             & \it disjoint & \it nested & \it halving & &  \\
\it general problem & \yes & \yes & \yes & NP-hard \citep{gabow} & \includegraphics[width=10em]{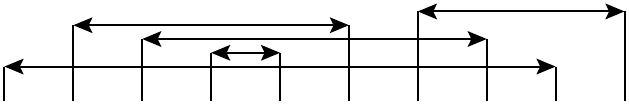} \\
\it overlapping structure & \no & \yes & \yes & NP-hard \citep{kp} & \includegraphics[width=10em]{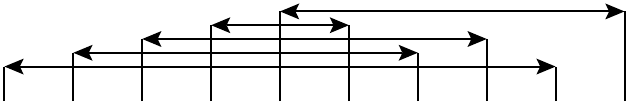} \\
\it ordered & \yes & \no & \yes & NP-hard [new] & \includegraphics[width=10em]{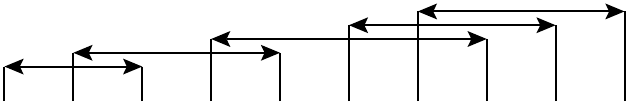} \\
\noalign{\medskip}
\it well-parenthesized & \yes & \yes & \no & $O(n^3)$ \citep{kp}, $O(n^\omega)$ [new] & \includegraphics[width=10em]{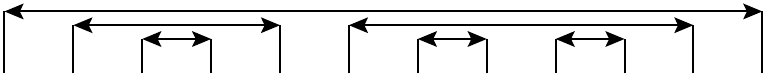} \\
\it halving & \no & \no & \yes & $O(n^5)$ \citep{kp}, $O(n^{\omega+1})$ [new] & \includegraphics[width=10em]{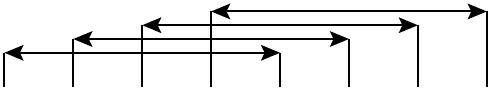} \\
\it nested & \no & \yes & \no & $O(nm)$ \citep{chen}, $O(n^\omega)$ [new] & \includegraphics[width=10em]{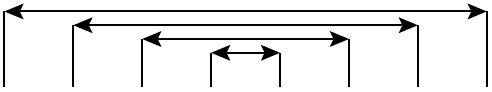} \\
\noalign{\medskip}
\it disjoint & \yes & \no & \no & $O(n+m)$ [trivial] & \includegraphics[width=10em]{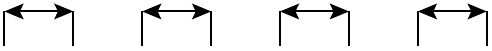}
\end{tabular}
}
\end{table*}

\begin{enumerate}
\item \emph{general case} -- there are no constraints on the positions of pairs;
\item \emph{overlapping structure}\footnote{note that this special case is refered to
      as \emph{halving structure} by \citet{kp}; we reserve the term ``halving"
      for sets where every two pairs halve each other}
      -- every two forbidden pairs overlap (they may be nested or halving, but not disjoint);
      as a consequence, 
      $f_1\prec f_2\prec \cdots\prec f_k \;\prec\; f_{\sigma(1)}'\prec f_{\sigma(2)}'\prec\cdots\prec f_{\sigma(k)}'$
      for some permutation $\sigma$;
\item \emph{ordered} -- there may be disjoint and halving pairs, but no two forbidden pairs are
      nested; as a consequence $f_1\prec f_2\prec\cdots\prec f_k$ and $f_1'\prec f_2'\prec\cdots\prec f_k'$;
\item \emph{well-parenthesized} -- there may be disjoint and nested pairs, but no two
      pairs are halving; this case deserves its name since if we write $\lp_i$ and $\rp_i$
      for the $i$-th pair, we get a well-parenthesized sequence;
\item \emph{halving} -- every two pairs halve each other;
      $f_1\prec f_2\prec\cdots\prec f_k \;\prec\; f_1'\prec f_2'\prec\cdots\prec f_k'$;
\item \emph{nested} -- there are only nested pairs, i.e., the vertices in forbidden pairs are ordered
      $f_1\prec f_2\prec \ldots\prec f_k \;\prec\; f_k'\prec \cdots\prec f_2'\prec f_1'$;
      this is a special case of the well-parenthesized case.
\end{enumerate}

The previous work and our own results are summarized in Table~\ref{tab:pafp}.

For completeness and as a warm-up, we include our own proof of NP-hardness of the PAFP problem
in the general and overlapping case. This proof is also simpler than the one given by \citet{kp}.

\begin{thm}\label{thm:pafp-np}
The PAFP problem is NP-hard, even when the set of forbidden pairs has overlapping structure.
\end{thm}

\begin{proof}
By reduction from 3-SAT: Let $\phi=\bigwedge_{1\leq i\leq n} \phi_i$ be a formula
over $m$ variables $x_1,\ldots,x_m$, with $n$ clauses $\phi_i= (\ell_{i,1}\vee \ell_{i,2}\vee \ell_{i,3})$,
where each literal $\ell_{i,j}$ is either $x_k$ or $\lnot x_k$. We will construct graph
$G$ and a set of forbidden pairs $F$ such that there is an $s$--$t$ path avoiding pairs in $F$
if and only if $\phi$ is satisfiable.

\begin{figure}[htp]
  \centering\includegraphics[scale=0.8]{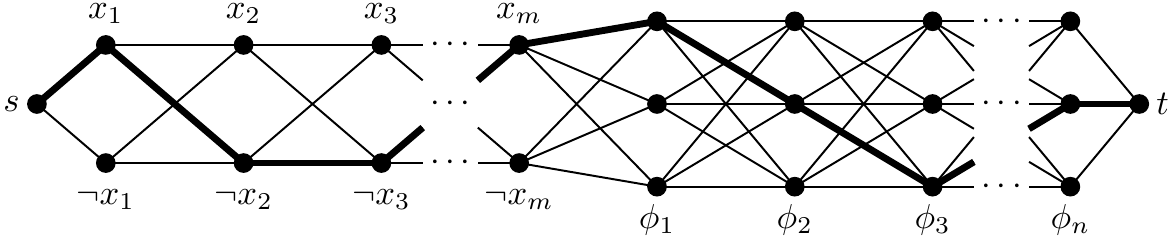}
  \caption{Input for the PAFP problem for the formula $\phi_1\land\phi_2\land\cdots\land\phi_n$.
           All edges are directed from left to right.}\label{fig:ppup-np}
\end{figure}

$G$ consists of two parts: The first part contains a vertex for each variable $x_k$ and its
negation $\lnot{x_k}$ (see Fig.~\ref{fig:ppup-np}). A path traversing this first part
corresponds to a truth assignment of variables where the visited vertices are true.
The second part contains a vertex for each literal $\ell_{i,j}$ (see Fig.~\ref{fig:ppup-np}).
Forbidden pairs connecting every literal from the first part to every occurence of its negation
in the second part of $G$ will ensure that we can only go through ``true'' vertices. Thus an $s$--$t$
path avoiding $F$ exists if and only if every clause is satisfied. Since every forbidden pair
starts in the first part and ends in the second part, all pairs overlap.
\end{proof}
\section{Ordered forbidden pairs} \label{s:ordered}

In this section, we turn to a seemingly more restricted version of the PAFP problem,
allowing only disjoint and halving forbidden pairs. This special case has not been
studied before.

\begin{thm}\label{thm:ord-np}
The PAFP problem is NP-hard, even when the set of forbidden pairs is ordered.
\end{thm}
\begin{proof}
We will prove the claim by a reduction from 3-SAT. Let $\phi$ be a
Boolean formula over $m$ variables $x_1,\ldots,x_m$, which is a
conjunction of $n$ clauses $\phi_1\wedge \cdots \wedge \phi_n$, where
$\phi_i= (\ell_{i,1}\vee \ell_{i,2}\vee \ell_{i,3})$ and each literal
$\ell_{i,j}$ is either $x_k$ or $\lnot x_k$. We will construct graph
$G$ with a linear order $\prec$ on its vertices and an ordered set
of forbidden pairs $F$ such that there is an $s$--$t$ path avoiding pairs in $F$
if and only if $\phi$ is satisfiable.

Graph $G$ consists of several blocks $B$ and $B_\ell$ of $2m$ vertices
shown in Fig.~\ref{fig:ord-block}, \subref{fig:ord-block2}. The blocks
are connected together as outlined in Fig.~\ref{fig:ord-graph}.
Any left-to-right path through the block $B$ naturally corresponds to
a truth assignment of the variables and, since $B_\ell$ has an isolated vertex $\lnot\ell$,
a path through block $B_\ell$ corresponds to an assignment where $\ell$ is true. 
A clause gadget consists of three such blocks, each corresponding to one literal.
Any $s$--$t$ path 
must pass through one of the three blocks, and thus choose an assignment
that satisfies the clause.
The forbidden pairs in $F$ will enforce that the assignment of
the variables is the same in all blocks. This is done by adding a
forbidden pair 
between all literals $\ell'$ in the $B_{\ell}$-blocks with their counterparts
$\lnot \ell'$ in the previous and the following $B$-block.

\begin{figure*}[tp]
  \centering
  \subfigure[Block $B$ -- vertices of this graph correspond to positive and
             negative literals; a path through this block corresponds to
             a truth assignment of the variables.]{
    \includegraphics[scale=0.8]{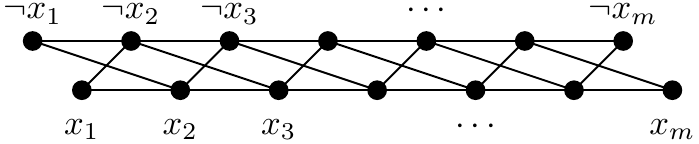}
    \label{fig:ord-block}
  }\qquad
  \subfigure[Block $B_\ell$ is similar to a $B$-block, except that the order of vertices
             is different and vertex $\lnot\ell$ is isolated. Thus, a path through
             $B_\ell$ corresponds to an assignment where $\ell$ is true.]{
    \includegraphics[scale=0.8]{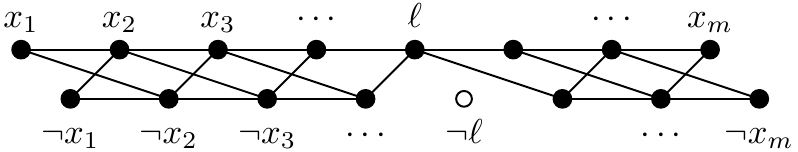}
    \label{fig:ord-block2}
  }
  \subfigure[Construction of $G$ from the blocks and zipped blocks corresponding to the clauses.
             Forbidden pairs enforce that the assignment of variables is the same in all blocks. 
            ]{
    \includegraphics{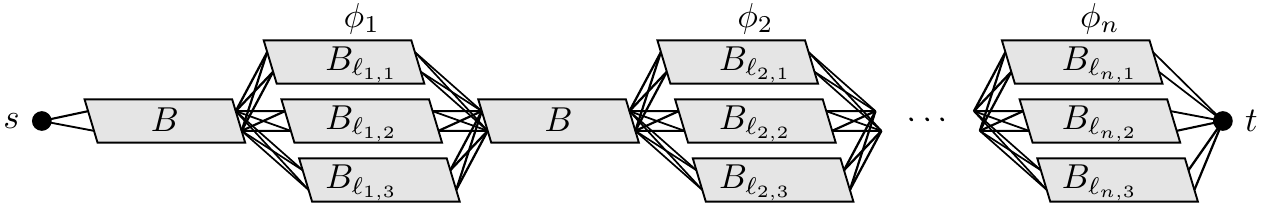}
    \label{fig:ord-graph}
  }
  \subfigure[An enlarged view of graph $G$ showing block $B$, the following blocks for clause $\phi_k$
    and the way they are connected by forbidden pairs. Note that no two forbidden pairs are nested.]{
    \includegraphics[scale=0.7]{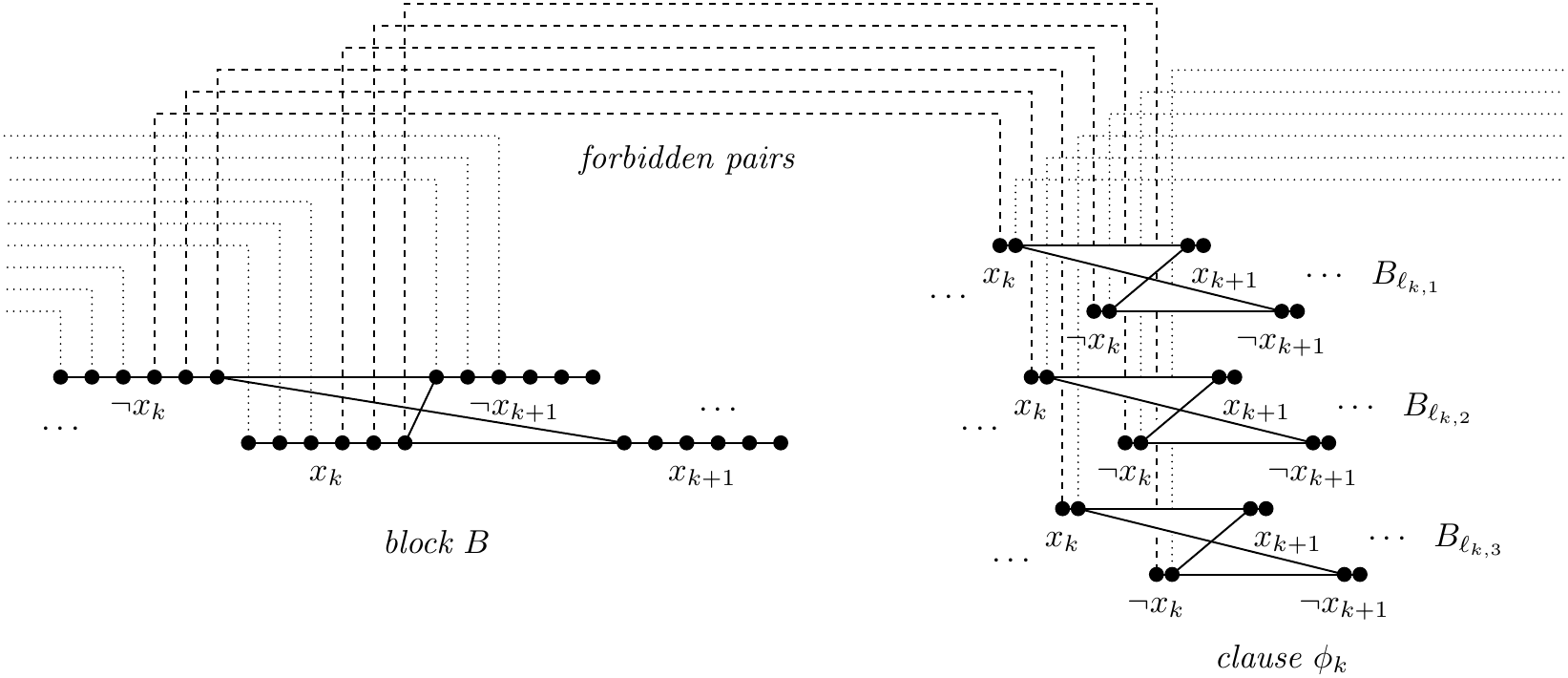}
    \label{fig:ordering}
  }
\caption{Construction of the graph $G$ for a 3-SAT formula $\phi$. All edges are directed from left to right.}
\end{figure*}

The order of literals in a $B$-block is $\lnot x_1\prec x_1\prec\lnot x_2
\prec\cdots\prec x_m$, while the order in a $B_\ell$-block is $x_1\prec\lnot x_1\prec x_2
\prec\cdots\prec\lnot x_m$. 
Let $v_1^i\prec v_2^i\prec v_3^i\prec\cdots$ be the order of vertices in graph $G^i$.
A \emph{zipping} operation takes graphs $G^1,G^2,G^3$ 
and produces a new graph $G^1\cup G^2\cup G^3$ with vertices
ordered $v_1^1\prec v_1^2\prec v_1^3\prec v_2^1\prec v_2^2\prec v_2^3\prec\cdots$.
The clause gadgets are produced by zipping the three blocks corresponding to their literals.
If we do not allow multiple forbidden pairs starting or ending in the same vertex, we
can substitute vertices in $G$ for short paths as in Fig.~\ref{fig:ordering}.
It is easy to check that under such linear order, no two pairs in $F$ are nested.
\end{proof}

\section{Well-parenthesized forbidden pairs} \label{s:paren}

The first polynomial algorithm for the PAFP problem with well-parenthesized forbidden
pairs was given by \citet{kp}. Their algorithm uses three rules for reducing the input graph:
\begin{enumerate}
\item \emph{contraction of a vertex} -- if $v$ does not appear in any forbidden pair,
      remove it and add a direct edge $(u,w)$ for every pair of edges $(u,v)$, $(v,w)$;
\item \emph{removal of an edge} -- if edge $e\in E\cap F$ joins two vertices that make
      up a forbidden pair, remove $e$ from $E$;
\item \emph{removal of a forbidden pair} -- if $(u,v)\in F$ is a forbidden pair,
      but there is no path from $u$ to $v$, remove $(u,v)$ from $F$.
\end{enumerate}
These three rules are alternately applied to the input graph until we end up
with vertices $s$ and $t$ only -- either joined by an edge or disconnected
-- which is a trivial problem.

A simple implementation of this approach gives an $O(n^2m)$ algorithm.
Using fast matrix multiplication, the time complexity can be reduced to $O(n^{\omega+1})\approx O(n^{3.373})$
and using a dynamic data structure for ``finding paths and deleting edges
in directed acyclic graphs" by \citet{Italiano}, it can be reduced still to
$O(n^3)$.

\begin{figure*}[ht] \centering
  \includegraphics[scale=0.8]{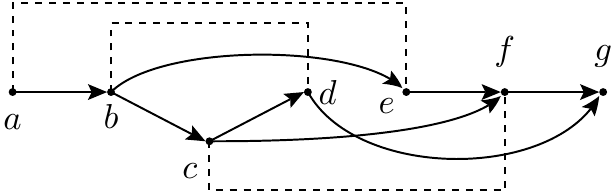}
  \caption{Example of a PAFP instance with hierarchical structure which has no well-parenthesized linear ordering:
  Since $b$ must be between $a$ and $e$, the only hope is to put $d$ before $e$ so that the pair $\{b,d\}$ is nested
  in $\{a,e\}$, but then the order must be $a,b,c,d,e,f,g$ and $\{c,f\}$ halves the other pairs.} \label{fig:counterex}
\end{figure*}

In fact, this algorithm does not need a topological ordering of vertices
and solves the PAFP problem for a larger class of instances having
a so called \emph{hierarchical structure}: if $\{u,v\}, \{x,y\}\in F$,
there is no path $u\to x \to v \to y$. In other words, no path contains
two halving pairs. Note that there are instances with hierarchical structure
such that no linear ordering is well-parenthesized (see Fig.~\ref{fig:counterex}).

Here we show that the linear ordering helps. 
First, we describe our own cubic algorithm, which is simple, does not use any advanced data structures
and moreover, can be easily extended to solve more general problems such as
\begin{itemize}
\item find an $s$--$t$ path passing the minimum number of forbidden pairs or
\item given a graph where all edges have scores and there are bonuses
      or penalties for some (well-parenthesized) pairs of vertices,
      find an $s$--$t$ path with maximum score (a problem motivated by an application in gene finding \citep{pcr}).
\end{itemize}
It seems unlikely that these problems can be solved using the former approach (because of rule 2).

After that, we show that our algorithm can be improved using the Valiant's technique
and fast matrix multiplication algorithms \citep{valiant,ukelson}
or the Four-Russians technique \citep{4russians}.
Note that the reduction to matrix multiplication is not only of theoretical interest,
since there are fast and practical hardware-based solutions for multiplying two matrices \citep{mm3, mm4}.

\goodbreak
\begin{thm}\label{thm:paren-pafp}
The PAFP problem with well-parenthesized forbidden pairs can be solved in $O(n^3)$ time.
\end{thm}

\begin{proof}
We first modify the input graph so that no two forbidden pairs start or end in the same vertex. 
Let $P[u,v]$ be true if a safe $u$--$v$ path exists, and let $J[u,v]$ be true if there is
a forbidden pair $(q,v)\in F$, $u\prec q \prec v$, and there is a safe $u$--$v$ path such that
the first edge jumps over $q$. 

The values of $P$ and $J$ can be found by dynamic programming:
It is easy to compute $J[u,v]$ (if we already know $P[w,v]$ for all $u\prec w\preceq v$)
by inspecting the neighbours of $u$. Conversely, we can also compute $P[u,v]$ efficiently
using the table $J$:
If no forbidden pair ends in $v$ or vertex $u$ is ``inside'' the forbidden pair $(q,v)\in F$, 
we just search the neighbours of $v$ for a vertex that could be penultimate on the $u$--$v$ path.
Otherwise, let $(q,v)\in F$ be a forbidden pair such that $u\prec q\prec v$.
Suppose that a safe $u$--$v$ path exists and let $w$ be the last vertex on this path before $q$.
Then $P[u,w]$ and $J[w,v]$ are both true.
Conversely, if $P[u,w]$ and $J[w,v]$ are true for some $w\prec q$, by concatenating the corresponding paths,
we get a safe $u$--$v$ path: The path obviously avoids all forbidden pairs before or after $q$
(from the definition of $P[u,w]$ and $J[w,v]$), and there are no forbidden pairs halving $(q,v)$.

Thus, $P[s,t]$ can be computed in cubic time using the following two recurrences:
\begin{numcases}{J[u,v]=}
\textstyle \bigvee_{(u,w)\in E,\,q\prec w} P[w,v] & \text{if $u\prec q$ and $(q,v)\in F$ is a forbidden pair}\\
\rlap{\it undefined}\hphantom{\textstyle \lor \bigvee_{u\preceq w\prec p} (P[u,w]\land J[w,v])} & \text{otherwise} \nonumber
\end{numcases}
\begin{numcases}{P[u,v]=}
\mathit{true}      & \rlap{\text{if $u=v$}}\hphantom{{if $u\prec q$ and $(q,v)\in F$ is a forbidden pair}} \nonumber\\
\mathit{false}     & \text{if $(u,v)\in F$ is a forbidden pair} \nonumber\\
\textstyle \bigvee_{u\preceq w\prec v,\ (w,v)\in E} P[u,w] & \text{if no forbidden pair ends in $v$ or $(q,v)\in F$ for $q\prec u$}\\
\textstyle \bigvee_{u\preceq w\prec q} (P[u,w]\land J[w,v]) & \text{if $(q,v)$ is a forbidden pair, $u\prec q\prec v$}
\end{numcases}
Obviously, each $J[u,v]$ and $P[u,v]$ can be computed in linear time, so the algorithm runs in $O(n^3)$.
\end{proof}

This algorithm can be further improved to $O(n^\omega)$ time by using fast boolean matrix multiplication.
The proof is actually simple 
thanks to the work of \citet{ukelson} that simplified and generalized the Valiant's technique \citep{valiant}.
They introduce a generic problem called \emph{Inside Vector Multiplication Template} (VMT) which can be solved
in subcubic time. A problem is considered an Inside VMT problem if it fulfills the following requirements:
\begin{enumerate}
\item The goal of the problem is to compute for every $i,j$ a series of inside properties
      $\beta_{i,j}^1,\beta_{i,j}^2,\ldots,\beta_{i,j}^K$.
\item Let $1\leq  k \leq K$, and let $\mu_{i,j}^k$ be a result of a vector multiplication
      of the form $\mu_{i,j}^k = \bigoplus_{q\in(i,j)}\left(\beta_{i,q}^{k'}\otimes\beta_{q,j}^{k''}\right)$,
      for some $1\leq k', k''\leq  K$. Assume that the following values are available: $\mu_{i,j}^k$,
      all values $\beta_{i',j'}^{k'}$ for $1 \leq  k' \leq  K$ and $(i',j')\subsetneq (i,j)$
      and all values $\beta_{i,j}^{k'}$ for $1 \leq  k' < k$.
      Then, $\beta_{i,j}^k$ can be computed in $o(n)$ time.
\item In the multiplication variant that is used for computing $\mu_{i,j}^k$, the $\oplus$ operation
      is associative, and the domain of elements contains a zero element. In addition, there is a matrix
      multiplication algorithm for this multiplication variant, whose running time $M(n)$ over two
      $n\times n$ matrices satisfies $M(n) = o(n^3)$.
\end{enumerate}

\begin{thm}[\citet{ukelson}]
For every Inside VMT problem there is an algorithm whose running time is $o(n^3)$.
In particular,
let $M(n)$ be the complexity of the matrix multiplication used and suppose
that $\beta_{i,j}^k$ can be computed in $\Theta(1)$ time in item 2 of the definition above.
Then the time complexity is $\Theta(M(n)\log n)$, if $M(n)=O(n^2\log^k n)$;
and $\Theta(M(n))$, if $M(n)=\Omega(n^{2+\varepsilon})$ for $\varepsilon>0$
and $4M(n/2)\leq d\cdot M(n)$ for some $d<1$ and sufficiently large $n$.
\end{thm}

\def\plbn{P^{\,\lceil\null\bullet}}
\def\plb{\plbn_{w,v}}
\def\pbln{P^{\,\bullet\null\lceil}}
\def\pbl{\pbln_{w,v}}

\begin{cor}
The PAFP problem with well-parenthesized forbidden pairs can be solved in $O(n^\omega)$ time,
where $2<\omega<2.3727$ is the exponent in the complexity of the boolean matrix multiplication.
\end{cor}

\begin{proof}
We formulate our solution from Theorem~\ref{thm:paren-pafp} as an Inside VMT problem.
The goal is to compute inside properties $A, J, \alpha, \beta, P, \plbn$, and $\pbln$.
Properties $J_{u,v}$ and $P_{u,v}$ correspond to the dynamic programming tables from
the proof of Theorem~\ref{thm:paren-pafp}, other properties are auxilliary. Property
$A$ is the adjacency matrix of graph $G$ and it is constant ($A_{u,v} = 1$ if and only if $(u,v)\in E$).
Properties $\alpha, \beta$ are used to store the partial results from cases (2) and (3)
in the computation of $P[u,v]$. Finally, the auxiliary properties $\plbn$ and $\pbln$
can be computed from $P$ in constant time and are defined as follows:
\begin{eqnarray*}
&& \plb = P_{w,v} \land (q \prec w) \quad\hbox{if $(q,v)\in F$, else \emph{false}} \\
&& \pbl = P_{w,v} \land (w \prec q) \quad\hbox{if $(q,v)\in F$, else \emph{false}} 
\end{eqnarray*}
Now we can rewrite the computation of $J_{u,v}$ and $P_{u,v}$ using boolean vector multiplication
as follows: 
\begin{align*}
\textstyle \bigvee_{(u,w)\in E,\,q\prec w} P[w,v] & \qquad\leadsto\qquad 
       J_{u,v} = \textstyle \bigoplus_{w\in(u,v)}(A_{u,w}\otimes \plb)    \tag{1'}\\
\textstyle \bigvee_{u\preceq w\preceq v,\ (w,v)\in E} P[u,w] & \qquad\leadsto\qquad 
      \alpha_{u,v} = \textstyle \bigoplus_{w\in(u,v)}(P_{u,w}\otimes A_{w,v}) \tag{2'}\\
\textstyle \bigvee_{u\preceq w\prec q} (P[u,w]\land J[w,v]) & \qquad\leadsto\qquad
      \beta_{u,v} = \textstyle \bigoplus_{w\in(u,v)}(\pbl\otimes J_{w,v})    \tag{3'}
\end{align*}
Property $P_{u,v}$ can be computed from $\alpha_{u,v}$ and $\beta_{u,v}$ in constant time.
\end{proof}
\section{The other cases and concluding remarks} \label{s:concl}

Note that the $O(n^\omega)$ algorithm for well-parenthesized forbidden pairs
also improves upon the result by \citet{chen} for the nested case, when the input graph is dense.
It remains an open problem whether there is a more efficient algorithm for the nested case.

An $O(n^{\omega+1})$ time algorithm for halving forbidden pairs is achieved by a refined version
of the algorithm given by \citet{kp}. Recall that in this case, the input graph $G$ consists of
two parts: all the forbidden pairs start in the first part, and end in the second part in the same order.
Let us denote the vertices in the first part $s\prec x_1\prec\cdots\prec x_n$ and vertices in the second
part $y_1\prec\cdots\prec y_n\prec t$, where $\{x_i,y_i\}$ are forbidden pairs.
We may assume that all vertices are accessible from $s$ and that $t$ is accessible from every vertex.

If there is a direct edge from $s$ to the second part or if there is an edge from the first part to $t$,
a safe $s$--$t$ path exists trivially. Otherwise, we reduce the halving case to $n$ instances of the nested case.
There will be a safe $s$--$t$ path in $G$ if and only if there is a safe $s$--$t'$ path in at least one of the produced instances.

First, remove all the $(x_i,y_j)$ edges, add a new terminal vertex $t'$, and reverse the direction of all edges
in the second part of $G$. Note that in this new order, $s\prec x_1\prec\cdots\prec x_n\prec t\prec y_n\prec\cdots\prec y_1\prec t'$,
the forbidden pairs are nested.
The $k$-th instance is obtained by adding edges $(x_k,t)$ and $(y_\ell,t')$ for each edge $(x_k,y_\ell)$, so
there is a safe $s$--$t$ path $s,\ldots,x_k, y_\ell,\ldots,t$ in the original graph $G$ if and only if
there is a safe $s$--$t'$ path $s,\ldots x_k, t,\ldots, y_\ell, t'$ in the new graph.

It remains an open problem whether a more efficient algorithm exists.

\subsubsection*{Acknowledgements.}
The autor would like to thank Bro\v na Brejov\' a for many constructive comments.
The research of Jakub Kov\'a\v c is supported by 
APVV grant SK-CN-0007-09,
Marie Curie Fellowship IRG-231025 to Dr.\ Bro\v na Brejov\' a,
Comenius University grant UK/121/2011, and
by National Scholarship Programme (SAIA), Slovak Republic.
Preliminary version of this work appeared in \citet{pcr}.

\bibliographystyle{elsarticle-num-names}
\bibliography{main}

\begin{thebibliography}{17}
\providecommand{\natexlab}[1]{#1}
\providecommand{\url}[1]{\texttt{#1}}
\providecommand{\urlprefix}{URL }
\expandafter\ifx\csname urlstyle\endcsname\relax
  \providecommand{\doi}[1]{doi:\discretionary{}{}{}#1}\else
  \providecommand{\doi}[1]{doi:\discretionary{}{}{}\begingroup
  \urlstyle{rm}\url{#1}\endgroup}\fi
\providecommand{\bibinfo}[2]{#2}

\bibitem[{Krause et~al.(1973)Krause, Smith, and Goodwin}]{krause}
\bibinfo{author}{K.~Krause}, \bibinfo{author}{R.~Smith},
  \bibinfo{author}{M.~Goodwin}, \bibinfo{title}{Optimal software test planning
  through automated network analysis}, in: \bibinfo{booktitle}{Proc.\ 1973 IEEE
  Symp.\ on Computer Software Reliability}, \bibinfo{pages}{18--22},
  \bibinfo{year}{1973}.

\bibitem[{Srimani and Sinha(1982)}]{srimani}
\bibinfo{author}{P.~K. Srimani}, \bibinfo{author}{B.~P. Sinha},
  \bibinfo{title}{Impossible pair constrained test path generation in a
  program}, \bibinfo{journal}{Inf. Sci.}
  \bibinfo{volume}{28}~(\bibinfo{number}{2}) (\bibinfo{year}{1982})
  \bibinfo{pages}{87--103}.

\bibitem[{Gabow et~al.(1976)Gabow, Maheswari, and Osterweil}]{gabow}
\bibinfo{author}{H.~N. Gabow}, \bibinfo{author}{S.~N. Maheswari},
  \bibinfo{author}{L.~J. Osterweil}, \bibinfo{title}{On Two Problems in the
  Generation of Program Test Paths}, \bibinfo{journal}{IEEE Trans. Software
  Eng.} \bibinfo{volume}{2}~(\bibinfo{number}{3}) (\bibinfo{year}{1976})
  \bibinfo{pages}{227--231}.

\bibitem[{Chen et~al.(2001)Chen, Kao, Tepel, Rush, and Church}]{chen}
\bibinfo{author}{T.~Chen}, \bibinfo{author}{M.-Y. Kao},
  \bibinfo{author}{M.~Tepel}, \bibinfo{author}{J.~Rush}, \bibinfo{author}{G.~M.
  Church}, \bibinfo{title}{A Dynamic Programming Approach to De Novo Peptide
  Sequencing via Tandem Mass Spectrometry}, \bibinfo{journal}{J. Comput. Biol.}
  \bibinfo{volume}{8}~(\bibinfo{number}{3}) (\bibinfo{year}{2001})
  \bibinfo{pages}{325--337}.

\bibitem[{Kov{\'a}{\v c} et~al.(2009)Kov{\'a}{\v c}, Vina{\v r}, and
  Brejov{\'a}}]{pcr}
\bibinfo{author}{J.~Kov{\'a}{\v c}}, \bibinfo{author}{T.~Vina{\v r}},
  \bibinfo{author}{B.~Brejov{\'a}}, \bibinfo{title}{Predicting Gene Structures
  from Multiple RT-PCR Tests}, in: \bibinfo{editor}{S.~Salzberg},
  \bibinfo{editor}{T.~Warnow} (Eds.), \bibinfo{booktitle}{WABI}, vol.
  \bibinfo{volume}{5724} of \emph{\bibinfo{series}{Lecture Notes in Computer
  Science}}, \bibinfo{publisher}{Springer}, ISBN
  \bibinfo{isbn}{978-3-642-04240-9}, \bibinfo{pages}{181--193},
  \bibinfo{year}{2009}.

\bibitem[{Yinnone(1997)}]{yinnone}
\bibinfo{author}{H.~Yinnone}, \bibinfo{title}{On Paths Avoiding Forbidden Pairs
  of Vertices in a Graph}, \bibinfo{journal}{Discrete Appl. Math.}
  \bibinfo{volume}{74}~(\bibinfo{number}{1}) (\bibinfo{year}{1997})
  \bibinfo{pages}{85--92}.

\bibitem[{Kolman and Pangr{\'a}c(2009)}]{kp}
\bibinfo{author}{P.~Kolman}, \bibinfo{author}{O.~Pangr{\'a}c},
  \bibinfo{title}{On the complexity of paths avoiding forbidden pairs},
  \bibinfo{journal}{Discrete Appl. Math.}
  \bibinfo{volume}{157}~(\bibinfo{number}{13}) (\bibinfo{year}{2009})
  \bibinfo{pages}{2871--2876}.

\bibitem[{Hajiaghayi et~al.(2010)Hajiaghayi, Khandekar, Kortsarz, and
  Mestre}]{checkpoint}
\bibinfo{author}{M.~Hajiaghayi}, \bibinfo{author}{R.~Khandekar},
  \bibinfo{author}{G.~Kortsarz}, \bibinfo{author}{J.~Mestre},
  \bibinfo{title}{The Checkpoint Problem}, in: \bibinfo{editor}{M.~J. Serna},
  \bibinfo{editor}{R.~Shaltiel}, \bibinfo{editor}{K.~Jansen},
  \bibinfo{editor}{J.~D.~P. Rolim} (Eds.), \bibinfo{booktitle}{APPROX-RANDOM},
  vol. \bibinfo{volume}{6302} of \emph{\bibinfo{series}{Lecture Notes in
  Computer Science}}, \bibinfo{publisher}{Springer}, ISBN
  \bibinfo{isbn}{978-3-642-15368-6}, \bibinfo{pages}{219--231},
  \bibinfo{year}{2010}.

\bibitem[{Bodlaender et~al.(2011)Bodlaender, Jansen, and Kratsch}]{w1}
\bibinfo{author}{H.~L. Bodlaender}, \bibinfo{author}{B.~M.~P. Jansen},
  \bibinfo{author}{S.~Kratsch}, \bibinfo{title}{Kernel Bounds for Path and
  Cycle Problems}, in: \bibinfo{booktitle}{Proc. of the 6th International
  symposium on Parameterized and Exact Computation (IPEC)},
  \bibinfo{year}{2011}.

\bibitem[{Coppersmith and Winograd(1990)}]{mm}
\bibinfo{author}{D.~Coppersmith}, \bibinfo{author}{S.~Winograd},
  \bibinfo{title}{Matrix Multiplication via Arithmetic Progressions},
  \bibinfo{journal}{J. Symb. Comput.} \bibinfo{volume}{9}~(\bibinfo{number}{3})
  (\bibinfo{year}{1990}) \bibinfo{pages}{251--280}.

\bibitem[{Williams(2011)}]{mm2}
\bibinfo{author}{V.~Williams}, \bibinfo{title}{Breaking the
  Coppersmith-Winograd barrier}, \bibinfo{year}{2011}.

\bibitem[{Italiano(1988)}]{Italiano}
\bibinfo{author}{G.~F. Italiano}, \bibinfo{title}{Finding Paths and Deleting
  Edges in Directed Acyclic Graphs}, \bibinfo{journal}{Inf. Process. Lett.}
  \bibinfo{volume}{28}~(\bibinfo{number}{1}) (\bibinfo{year}{1988})
  \bibinfo{pages}{5--11}.

\bibitem[{Valiant(1975)}]{valiant}
\bibinfo{author}{L.~G. Valiant}, \bibinfo{title}{General Context-Free
  Recognition in Less than Cubic Time}, \bibinfo{journal}{J. Comput. Syst.
  Sci.} \bibinfo{volume}{10}~(\bibinfo{number}{2}) (\bibinfo{year}{1975})
  \bibinfo{pages}{308--315}.

\bibitem[{Zakov et~al.(2011)Zakov, Tsur, and Ziv-Ukelson}]{ukelson}
\bibinfo{author}{S.~Zakov}, \bibinfo{author}{D.~Tsur},
  \bibinfo{author}{M.~Ziv-Ukelson}, \bibinfo{title}{Reducing the worst case
  running times of a family of RNA and CFG problems, using Valiant's approach},
  \bibinfo{journal}{Algorithms Mol. Biol.}
  \bibinfo{volume}{6}~(\bibinfo{number}{1}) (\bibinfo{year}{2011})
  \bibinfo{pages}{20}.

\bibitem[{Arlazarov et~al.(1970)Arlazarov, Dinic, Kronrod, and
  Faradzev}]{4russians}
\bibinfo{author}{V.~Arlazarov}, \bibinfo{author}{E.~Dinic},
  \bibinfo{author}{M.~Kronrod}, \bibinfo{author}{I.~Faradzev},
  \bibinfo{title}{On economic construction of the transitive closure of a
  directed graph}, in: \bibinfo{booktitle}{Soviet Math. Dokl.},
  vol.~\bibinfo{volume}{11}, \bibinfo{pages}{1209--1210}, \bibinfo{year}{1970}.

\bibitem[{Ryoo et~al.(2008)Ryoo, Rodrigues, Baghsorkhi, Stone, Kirk, and mei
  W.~Hwu}]{mm3}
\bibinfo{author}{S.~Ryoo}, \bibinfo{author}{C.~I. Rodrigues},
  \bibinfo{author}{S.~S. Baghsorkhi}, \bibinfo{author}{S.~S. Stone},
  \bibinfo{author}{D.~B. Kirk}, \bibinfo{author}{W.~mei W.~Hwu},
  \bibinfo{title}{Optimization principles and application performance
  evaluation of a multithreaded GPU using CUDA}, in:
  \bibinfo{editor}{S.~Chatterjee}, \bibinfo{editor}{M.~L. Scott} (Eds.),
  \bibinfo{booktitle}{PPOPP}, \bibinfo{publisher}{ACM}, ISBN
  \bibinfo{isbn}{978-1-59593-795-7}, \bibinfo{pages}{73--82},
  \bibinfo{year}{2008}.

\bibitem[{Volkov and Demmel(2008)}]{mm4}
\bibinfo{author}{V.~Volkov}, \bibinfo{author}{J.~Demmel},
  \bibinfo{title}{Benchmarking GPUs to tune dense linear algebra}, in:
  \bibinfo{booktitle}{SC}, \bibinfo{publisher}{IEEE/ACM}, ISBN
  \bibinfo{isbn}{978-1-4244-2835-9}, \bibinfo{pages}{31}, \bibinfo{year}{2008}.

\end{thebibliography}
\end{document}